**ABSTRACT**

Recent instrumentation has demonstrated that the solar atmosphere supports omnipresent transverse waves, which could play a key role in energizing the solar corona. Large-scale studies are required in order to build up an understanding of the general properties of these transverse waves. To help facilitate this, we present an automated algorithm for identifying and tracking features in solar images and extracting the wave properties of any observed transverse oscillations. We test and calibrate our algorithm using a set of synthetic data which includes noise and rotational effects. The results indicate an accuracy of 1-2% for displacement amplitudes and 4-10% for wave periods and velocity amplitudes. We also apply the algorithm to data from the Atmospheric Imaging Assembly (AIA) on board the *Solar Dynamics Observatory* (SDO) and find good agreement with previous studies. Of note, we find that 35 – 41% of the observed plumes exhibit multiple wave signatures, which indicates either the superposition of waves or multiple independent wave packets observed at different times within a single structure. The automated methods described in this paper represent a significant improvement on the speed and quality of direct measurements of transverse waves within the solar atmosphere. This algorithm unlocks a wide range of statistical studies that were previously impractical.

## (1) INTRODUCTION

Two of the most persistent and elusive questions in solar and heliophysics are *"what heats the corona?"* and *"what accelerates the solar wind?"* Within the last four decades, numerous mechanisms and theories have been proposed, including (but not limited to) nanoflares (Gold, 1964; Parker 1972), electric potentials (Lemaire & Scherer 1971), Alfvén waves (Heyvaerts & Priest 1983), ion-cyclotron waves (Isenberg & Hollweg 1982), and magnetic reconnection

(Crooker et al. 2002; Fisk 2003). Magnetohydrodynamic waves, particularly Alfvén(ic) waves, have been a topic of intense study (e.g., Cranmer & Van Ballegooijen 2005; Verdini & Velli, 2007; van der Holst 2014), as they can simultaneously address both coronal heating and solar wind acceleration. While Alfvénic waves have been detected in the solar wind from in situ measurements since the 1970s (Belcher & Davis 1971), it is only recently that propagating transverse waves have been observed in the corona (Tomczyk et al. 2007; McIntosh et al 2011; Thurgood et al. 2014; Morton et al. 2015, 2016b).

Despite their critical role in theories and models, relatively few comprehensive and statistically rigorous studies of propagating transverse waves in the corona have been performed. There are two apparent reasons for this. First, it was not until the 2010 launch of the *Solar Dynamics Observatory* (SDO; Pesnell et al. 2012), that we had both the spatial and temporal resolution to observe and continuously measure the relatively small-scale transverse oscillations of coronal structures. Second, direct observations are typically time- and labor-intensive and require either manual validation of fit parameters or the averaging of many structures into a mean power spectrum. Therefore, most previous studies either analyzed only a handful of events (e.g. Aschwanden et al. 1999) or relied on indirect observational methods such as non-thermal broadening of spectral lines (Banerjee et al. 2009).

Despite the challenges, a number of different methods have been developed for analyzing waves in the corona. Techniques for measuring propagating intensity disturbances, which display translational motion parallel to the local magnetic field, have the greatest variety. Existing codes for detecting propagating disturbances utilize cross- and 2D coupled fitting methods (Yuan and Nakariakov 2012), the application of surfing transforms (Uritsky et al. 2013), wavelets (Krishna Prasad et al. 2011), and running difference images (Sheeley et al. 2014). In contrast, methods for detecting transverse waves are less diverse. The oscillations are identified using either visual inspection (e.g. Zimovets & Nakariakov 2015) or time-difference images (see Aschwanden & Schrijver 2011) and the wave parameters are most commonly measured by manually fitting a sinusoidal function to the peak intensity location. Such methodologies require considerable time and effort but nevertheless have been used effectively to analyze both damped and decay-less standing transverse waves in coronal loops (Anfinogentov et al. 2013; Nisticò et al. 2013; Pascoe et al. 2016). There has also been some work to measure the wave properties using wavelet transforms (Verwichte et al. 2004; Nisticò et al. 2014) but it is unclear if the codes may be used in a generalized or automated manner. Recently, an algorithm for "motion magnification" has been developed by Anfinogentov and Nakariakov (2016). This tool enhances the appearance of low-amplitude transverse oscillations so that they may be more easily visualized and measured by other wave analysis codes.

In this paper, we present an extension of the Northumbria University Wave Tracking (NUWT) code (Morton et al. 2016a), an automated algorithm for identifying and analyzing transverse waves in a series of images. The fundamental framework of NUWT was developed by Morton et al. (2013) and previous versions of the code were used in the plume studies of Thurgood et al. (2014) and Morton et al. (2015). In § 2 and § 3 we describe the basic operation of NUWT and use comparisons to synthetic datasets as a means of validation and estimating the accuracy of returned parameters. Then, in § 4, we apply NUWT to data obtained by the Atmospheric Imaging Assembly (AIA; Leman et al. 2012) on board *SDO* and present results from five different four-hour time periods. Finally, we discuss our findings in light of previous measurements and consider additional factors that may be significant. It is important to note here that, while our specific examples and initial applications use solar images from *SDO*/AIA, the methods of NUWT are sufficiently general that the code may be applied to any set of imaging data. The two primary requirements are that: (a) the instrument has sufficient spatial resolution and temporal cadence to observe the transverse wave motions of interest and (b) the waves are propagating along features that have intensities that are either bright local maxima or dark local minima.

## (2) METHODOLOGY OF THE NUWT CODE

Fundamentally, NUWT operates by extracting a virtual data slit from an input series of images, identifying and tracking "threads" of local extrema, and measuring any transverse wave behavior present. For simple slit geometries, this enables a fully automated analysis process in which the user need only supply a set of images and, optionally, modify a few run parameters. Alternatively, NUWT may operate as a semi-automated component of a larger, more complicated analysis involving data slits either manually processed by the user or extracted by a specialized program designed to identify the particular features of study.

There are six basic steps in the NUWT data processing pipeline:

1. Data acquisition and preprocessing

2. Slit Extraction

3. Feature Identification

4. Thread Tracking

5. Application of FFT

6. Filtering Waves and Calculating Observables

Additional details for steps 3 and 4 can be found in Morton et al. (2013) and Morton et al. (2014).

***Step 1:*** *Data acquisition and preprocessing.* First, the data must be acquired and all instrument-specific corrections and processing must be performed. For solar imaging data, such processes include rotating, rescaling, co-alignment, and de-spiking. Afterward, two optional preprocessing filters may be applied: an image sharpening filter (e.g., unsharp masking) to highlight fine-scale features and a temporal filter to help suppress random noise and frame-to-frame intensity variations. Fundamentally, these filters involve applying a boxcar average in, respectively, the spatial and temporal dimensions. The particular size of the filtering windows used should depend on the characteristic scale of the features being investigated and the cadence of the data. We have found that, in general, large values of noise, e.g., from cosmic-ray hits, negatively influence later steps in the processing. Hence, if the temporal filter is not used, we suggest some other effort to suppress noise values should be made.

***Step 2:*** *Slit extraction.* Next a virtual data slit is defined and a two-dimensional time-distance diagram is constructed by extracting the data values along the slit at each time step. These diagrams show the locations of bright (as well as dark) structures that cross the slit and whose motion is projected onto the observational plane. The intensity uncertainties for each value in the time-distance diagram are also extracted or estimated as part of this step, and should include expected contributions from standard sources, such as photon noise, dark current, etc. (e.g., Yuan & Nakariakov 2012, Morton et al. 2014). It is important that a reasonable estimate for data errors is given, as this influences the uncertainties on model parameter values that are calculated from fitting to the data at a later stage.

***Step 3:*** *Feature identification.* All local maxima are found in the time-distance diagram by comparing values to their $N$th nearest neighbors along the distance axis. The choice of $N$ will determine the minimum allowed distance between the detected structures. Setting $N$ too small yields noisy and spurious results, while setting $N$ too large will cause dim structures to be overlooked. By default, a value of $N = 5$ is chosen from experience with *SDO* data, corresponding approximately to a coronal loop width and the distance between neighboring loops (e.g., Aschwanden & Nightingale 2005; Brooks et al., 2013; Morton & McLaughlin 2013, Reale 2014; Scullion et al. 2014; Aschwanden & Peter 2017). This value, however, will likely need to be modified for higher resolution data. Next, the maxima are filtered by comparing the gradients on either side to a user-adjustable threshold. The threshold used should balance letting through segments of structures that have small spatial intensity gradients and minimizing random noise. For unsharp-masked data, a gradient threshold of 0.5 is sufficient in most cases. Again, the threshold will be data dependent and a process of trial and error should be undertaken to find a threshold value that provides the right balance. A complementary approach would be to use image classification techniques to suppress spurious peaks due to noise (e.g., majority analysis or sieving); however such techniques have yet to be tested in conjunction with NUWT. Finally, the sub-pixel location of each selected maxima is determined by fitting the nearby intensity values with a Gaussian model, using a non-linear least-squares fitting method (Markwardt 2009) which takes into account the intensity uncertainties. Fits that return central locations of the Gaussian that differ from the whole-pixel maxima location by more than 1.5 pixels are rejected and the program defaults to the original whole-pixel value.

*Step 4:* Thread tracking. The fourth step is to track the time evolution of each feature by connecting the maxima into "threads" – which are essentially time-series of each feature's displacement. This is performed using a nearest-neighbor method that scans a search box in space and time. The size of the search box can be adjusted to limit the maximum permissible transverse velocity, as well as the maximum number of permitted missing data points between maxima at different times. By default, the largest allowable data gap is 3 missing points and the maximum frame-to-frame displacement is 3 pixels (for AIA, this corresponds to a maximum velocity amplitude of ~100 km/s). Each thread is only permitted to have a single value at each time-step and a given peak cannot be assigned to more than one thread. As part of this step, threads that do not contain a minimum number of data points are rejected (20 data points is found to be reasonable cut-off). Furthermore, any threads with >35% of data points missing are also rejected.

*Step 5: Application of FFT*. The fast Fourier transform requires regularly sampled data without gaps (e.g. Munteanu et al. 2016). Therefore, we fill gaps within each thread using linear interpolation. Next we apply a split cosine bell windowing function to the time-series and, optionally, apply zero-padding. We then run the FFT method and correct the output power spectrum to account for signal lost due to windowing and zero-padding.

*Step 6: Filtering waves and calculating observables*. The final step of NUWT is to select the significant wave components of the FFT power spectrum. Although data from the *Coronal Multi-channel Polarimeter* (CoMP) indicates that the time-averaged behavior of coronal waves exhibits a power-law spectrum (Morton et al., 2016b), the individual wave packets observed by NUWT have near sinusoidal motion with potentially multiple superimposed wave packets. Therefore, we use a null hypothesis test based on a white-noise power spectrum, with the significance threshold calculated from the data using (Torrence & Compo 1998):

$$P_{white\ noise} = \frac{\sigma^2 \chi^2}{ND}$$

where $\sigma$ is the standard deviation of the time-series, $\chi^2$ is the cut-off value of the chi-square distribution at the selected significance level, $N$ is the number of points in the FFT spectrum, and

*D* is the degrees of freedom. By default, NUWT uses an adjusted significance level of 5% after applying the "Bonferroni correction" for multiple, simultaneous significance tests (See Armstrong 2014 and the references therein). All peaks with power greater than the significance threshold are identified as different waves propagating on the same structure. The wave displacement amplitudes are calculated from the power spectrum and the velocity amplitudes are calculated using the relation $v = 2\pi\xi f$, where $\xi$ is the displacement amplitude and *f* is the frequency of the wave. Results for threads composed of 35 - 50% data gaps are retained for diagnostic purposes but omitted from the calculation of summary statistics.

## (3) CALIBRATION AND VALIDATION OF NUWT

### (3.1) SYNTHETIC AIA DATA

Before we examine the results obtained with NUWT from observational data, we estimate the accuracy and fundamental limitations of the algorithm. To this end, we generated a set of synthetic time-distance diagrams containing structures undergoing oscillatory displacements. In the most basic case, we simulated 3,000 distinct structures in a 3,000 by 48,000 pixel box which represents a 600 arcsecond wide slit observed over 12 hours with a spatial resolution of 0.2 arcsec and a cadence of 0.9 seconds. For simplicity, we only generated one wave on each structure and confined the motions to the observational plane. Each structure was given a Gaussian cross-sectional intensity profile, where the Gaussian amplitude and width were held constant over the lifetime of the structure. However, the intensity amplitudes were randomly selected for each structure to represent features with different emission measures. The central locations of the Gaussian in each structure were shifted, with the locations in time defined by sinusoids of the form $y = \xi \cos(2\pi t f + \varphi)$, where $\xi$ is the displacement amplitude of the wave, *f* is the frequency, and $\varphi$ is the phase. The amplitudes and frequencies of the wave motions were randomly sampled from Gaussian distributions with known means and standard deviations. The number of cycles simulated for each structure was varied between 0.5 and 2.0 cycles and this, in turn, determined the total duration of each structure. Next, we generated all of the waves and randomly distributed them within the high-resolution simulation box. Then we degraded the spatial resolution to 0.6 arcseconds and the temporal resolution to 2.7-second-long exposures at a cadence of 12 seconds. These values were chosen to approximate the resolution and cadence of *SDO*/AIA images in the 171 Å band. Finally, we added a constant background intensity level with both artificial white-noise added and Poisson noise applied in-line with expected noise levels. Figure 1 shows an example section of the final synthetic time-distance diagram after the background and noise has been added. As one can see, some structures have peak intensities near the background level and are difficult to distinguish from the noise.

We note that this simulation setup is likely to mirror situations with a reasonably simple magnetic geometry, e.g., structures in the quiescent Sun, coronal holes. For active regions, the increased magnetic complexity can lead to structures crossing over each other during the time period of oscillations. Given that it is often the goal to analyze the time-evolution of a transverse wave in an active region over a number of cycles (e.g., to estimate damping coefficients from the amplitude envelope), the crossing of structures would lead to NUWT breaking apart threads and analyzing them individually. Therefore, additional care must be used when interpreting the NUWT results for such highly dynamic situations.

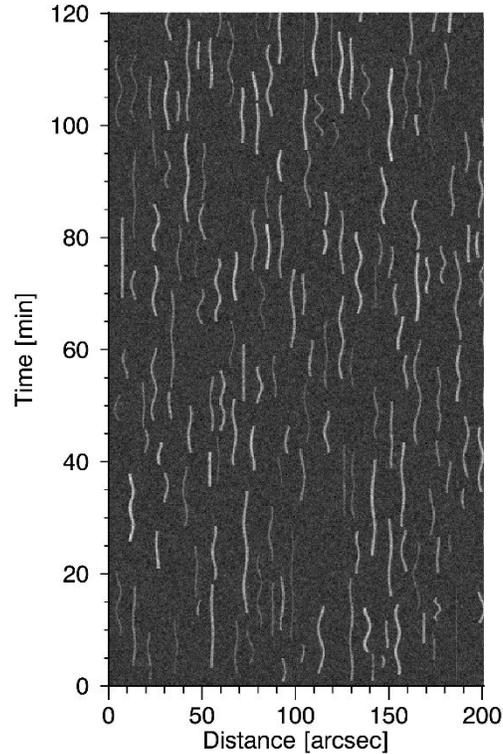

**Figure 1.** Example section of the synthetic time-distance diagram for the basic case of 3,000 waves oscillating in the plane of observation. The simulated structures were given a range of intensities relative to the background and both white- and Poisson noise have been applied.

We ran the synthetic AIA data through the NUWT algorithm and matched each thread detected with its corresponding input structure. NUWT found a total of 2,714 threads, which contained 2,933 out of the 3,000 input structures. Close inspection of the detected threads reveals that the allowance for gaps in thread observations resulted in 369 of the input structures being incorrectly appended to the ends of other threads. Conversely, interference from the added noise resulted in 56 input structures becoming split and returned by NUWT as 114 separate threads. The FFT method identified a single wave component for 2,415 (89.0%) threads, two wave components for 162 (6.0%) threads, and three or more wave components for 71 (2.6%) threads. These additional wave components are, of course, incorrect since only one wave was inputted for each thread. However, out of the 233 threads with multiple waves, 166 were caused by multiple input structures being combined into a single output thread. The remaining 67 multi-wave events are due to the addition of noise values to the ends of short threads. A total of 66 (2.4%) of the 2,714 detected threads were found to have no significant wave signal. This suggests that the false-positive detection rate of the NUWT is low (in this instance, no spurious threads or waves were composed entirely of noise values). However, there is some potential mixing of signals between different structures and from the combination of structures and noise.

We find that as thread length increases, the agreement between the input and output values becomes better. Generally poor results were obtained for threads with less than 20 data points. This is what is expected of an FFT-based method; more accurate results are returned for time-series with more data points since, in most cases, this corresponds to a greater number of observed oscillation cycles. Most of the 89 input structures that were missed by NUWT either had lengths shorter than 20 data points or were split by noise into segments with lengths of 20 or less. Figure 2 shows histograms of the displacement amplitudes, periods, and velocity amplitudes for all NUWT threads with more than 20 data points and with at least one identified wave. The red lines show the histograms of the input parameters for the corresponding simulated structures. The comparison between the mean and median values of each distribution indicates that the NUWT results are within 1.9% of the input displacement amplitudes, 7.2% of the input periods, and 6.4% of the input velocity amplitudes. Considering the limitations of windowed FFT methods, we believe that this is a reasonable level of accuracy to expect.

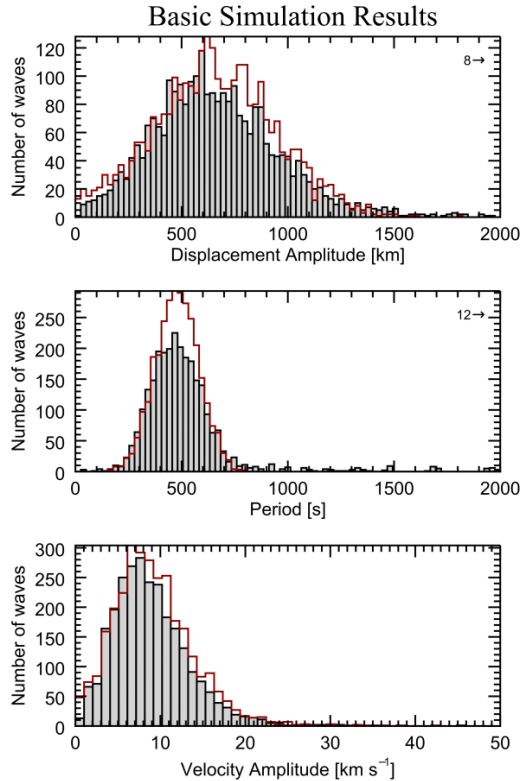

**Figure 2.** Histograms of the wave parameters found by NUWT (grey) and the associated input parameters (red) for the basic simulation. The log-normal means given by NUWT are within 1.9% of the input displacement amplitudes, 7.2% of the input periods, and 6.4% of the input velocity amplitudes.

As a test for NUWT's robustness against false-positive identification of wave motions, we generated a synthetic AIA image consisting entirely of noise values (no structures with any waves were inputted). Using the same parameters as the basic wave simulation, NUWT returned 40 false threads. However, after filtering for <35% data gaps, only 5 significant wave results were found. Therefore, we can conclude that the default NUWT values are sufficiently capable of rejecting results from pure background noise and are robust against potential false-positive thread identifications.

## (3.2) OUT-OF-PLANE WAVE COMPONENTS

Without additional information concerning the topology of the coronal magnetic field, we have no reason to assume that all waves are oscillating in the same plane. Therefore, we next generated a synthetic dataset in which the waves were allowed to have oscillations rotated at any angle relative to the observational plane, and we examined how well NUWT recovered the underlying (i.e., unrotated) distribution of wave properties. Inclusion of a rotation angle of $\theta$ will reduce the apparent amplitudes of transverse waves within the plane of observation. This was modeled by adding a factor of $cos(\theta)$ to the generated wave equations, where $\theta$ was picked from a uniform distribution between 0 and $+2\pi$. Therefore, analytically, the displacement amplitudes should be reduced by an average value of $\sqrt{2}$.

Similar to the basic synthetic case, we simulated 3,000 rotated waves in a 600 arcsecond by 4 hour time-distance diagram. Using a length threshold of 20 data points, NUWT found 2,744 threads, which corresponded to 2,930 of the input structures. 2,267 of the detected threads were matched one-to-one with the associated input structures while 333 threads were composed of multiple input structures, 137 threads corresponded to partial segments, and 7 threads were unpaired. Out of the 2,744 threads, 275 (10.0%) were found with no significant wave signals, 2,156 (78.6%) exhibited one wave, 231 (8.4%) were found with two waves, and 82 (3.0%) had three waves or more. Therefore, we conclude that the addition of a rotation angle reduces the total number of detected waves but does not significantly affect the proportion of threads with multiple waves.

The left-hand column of Figure 3 shows histograms of the unmodified wave parameters found by NUWT for simulated waves with arbitrary rotations. We note that the rotation of the waves relative to the observation plane does not greatly affect the identified periods; however, the distribution of displacement amplitudes are, as expected, shifted to lower values. We find that the sample log-normal mean amplitude is reduced by a factor of 1.407 which is very close to the analytical value. If we scale the NUWT amplitudes by a constant factor of $\sqrt{2}$, as illustrated in the right-hand column of Figure 3, we obtain sample means with similar accuracy as found in § 3.1. The scaled log-normal displacement amplitudes are within 0.5% of the input displacement amplitudes, 9.2% of the input periods, and 4.0% of the input velocity amplitudes. It should be noted that this scaling is only appropriate when calculating the bulk sample statistics. Individual results may still have a large difference between the input and returned values, as evidenced by the discrepancy between the distributions of the amplitude shown in Figure 3. We suggest that an appropriate correction can be applied to the measured distributions based on a Monte Carlo scheme (see, e.g. Morton & Swift, 2014 for an example of distribution corrections for exoplanets); however, this lies outside the focus of the current manuscript and will be addressed in future work.

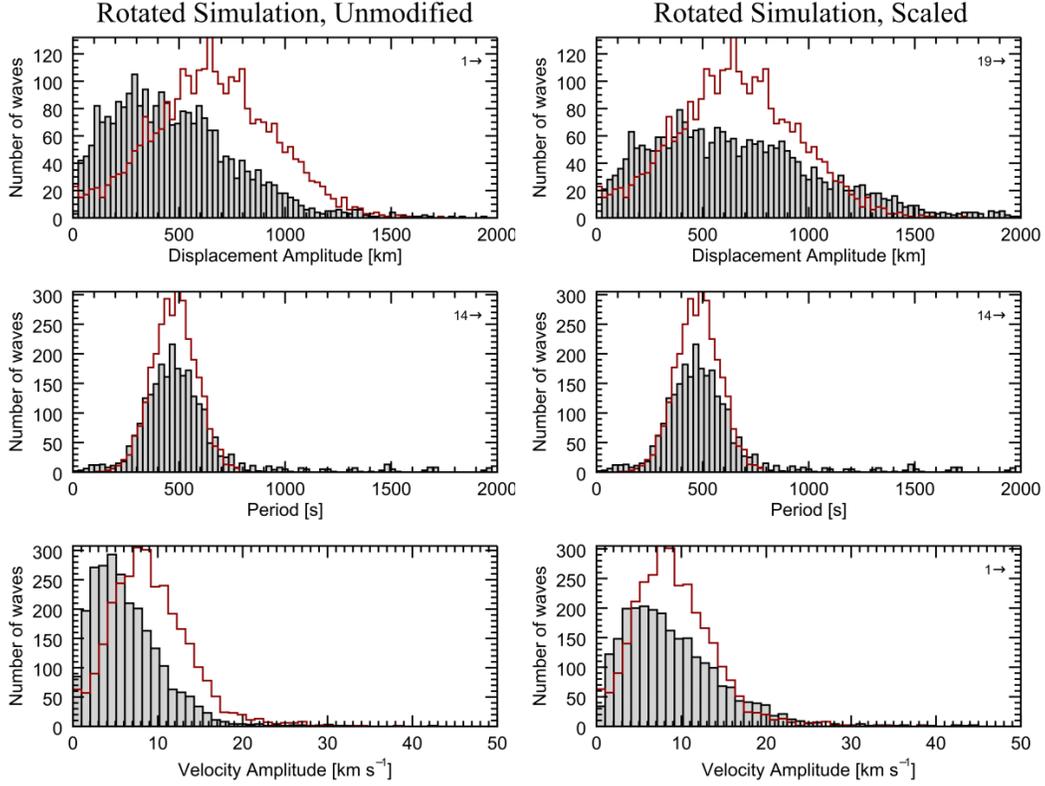

**Figure 3.** Wave parameters returned by NUWT (grey) compared to their input distributions (red) for the simulation with rotated structures. The left-hand side depicts the NUWT results without any modification or scaling. As expected, the observed amplitudes are reduced. The right-hand side shows the NUWT results scaled by a constant factor of $\sqrt{2}$. While there remains large differences in the distribution tails, the calculated bulk parameters better represent the input values.

## (4) APPLICATION TO *SDO* / AIA DATA

We begin investigations of the *SDO* data by focusing on features in regions of open magnetic field. Solar coronal plumes are faint ray-like features that fan outward from coronal holes in white-light and EUV images (see Wilhelm et al. 2011; Poletto 2015). Plumes are cooler and denser than the surrounding plasma with typical temperatures in the range of 0.7 – 1.2 MK and density enhancements of 2-3 (Ahmad & Withbroe 1977; Deforest et al. 1997; Del Zanna et al. 2003). Since plumes are bright relative to their environment, the motion and spectra of plumes have been used to detect and estimate the characteristics of waves in coronal holes (DeForest & Gurman 1998; McIntosh et al. 2011; Thurgood et al. 2014; Morton et al. 2015)

## (4.1) DATA SELECTION

Having established bounds on the accuracy of NUWT, we now apply it to data from the 171 Å channel of AIA. We analyzed three different four-hour time periods: 04:00–08:00 UT on 2010 May 23, 00:00–04:00 UT on 2010 August 6, and 16:00–20:00 UT on 2012 March 27. These dates and times were chosen to correspond with the time periods analyzed in previous studies of transverse waves in coronal plumes. The images in each dataset have a resolution of 0.5995" and a nominal cadence of 12 seconds. After processing the data to level 1.5 using the *aia_prep.pro* routine from the SolarSoft library, we apply a 6″ x 6″ unsharp mask and smooth the data over 3 time-steps to order to highlight small-scale features and suppress frame-to-frame intensity variations. Intensity errors were estimated using the methodology of Yuan & Nakariakov (2012) and the calibration parameters for the AIA 171 Å channel reported by Boerner et al. (2012). We then selected a total of five arc-shaped data slits. Two slits were located 15 Mm above the south polar limb on 2010 May 23, the first was above the south polar coronal hole (CH) and the second **was** within a quiet-sun region (QS). The third selected slit was 15 Mm above the north polar coronal hole on 2010 August 6. The final two slits were positioned at 7 Mm and 15 Mm within an open field (OF) region above the solar north pole on 2012 March 27. Each of the five slits span 10º of heliographic latitude (~200″) and are shown in Figure 4.

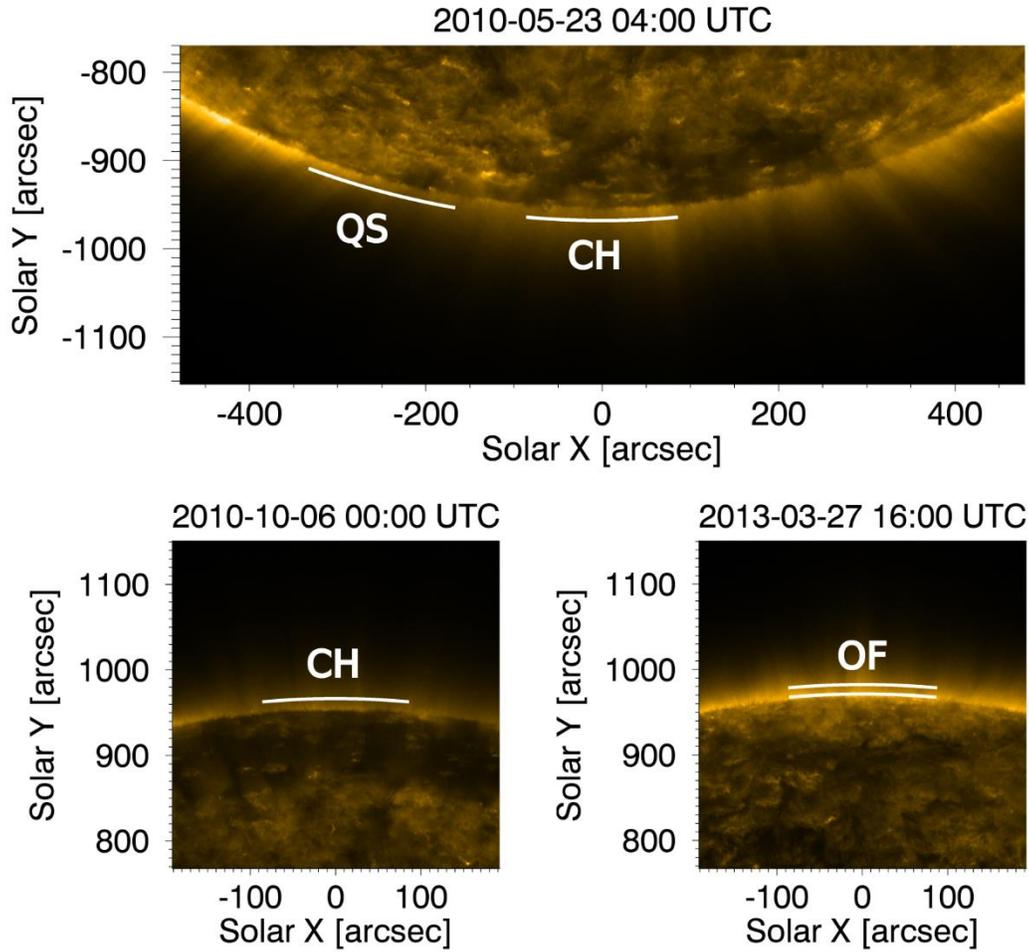

**Figure 4.** Locations of the five data slits selected for analysis. Each time period spans 4 hours of data were collected by the 171 Å channel of *SDO* / AIA. The extracted slits have lengths of ~200 arcsec and finite widths of 1.8 arcsec (3 pixels). Two slits were selected on 2010-05-23 (top), one 15 Mm above the south polar coronal hole, and another 15 Mm above a quiet-sun region. Another coronal hole slit was extracted from 15 Mm above the solar north pole on 2010-10-06 (lower left). Finally, two slits were selected on 2012-03-27 (lower right) at heights of 7 and 15 Mm above an open field region at the solar north pole.

Table 1 summarizes the basic NUWT results for all five data slits. Given the lessons learned in § 3.1, the minimum thread length threshold in each run was set to 20 data points and FFT padding was utilized. The number and distribution of events are similar between the data slits. Altogether, NUWT identified 2,470 threads with a total of 3,338 waves across all slits. Between 43.7% – 57.1% of threads exhibit a single wave, 25.4% – 29.2% have two superimposed waves, and 7.1% – 12.6% have three waves or more. No threads were found with five or more waves. Furthermore, 6.7% – 18.2% of the detected threads had no significant waves, although many of the waveless spectra have peaks just below the significance threshold, suggesting either underresolved wave motions or too strict of a threshold. The open field slit at 7 Mm on 2012 March 27 has both the lowest fraction of threads with no waves and the highest proportion of threads with one or two waves. The overall mean thread duration is 676 s (~56 data points).

**Table 1.** Number of threads and waves found by NUWT in coronal hole (CH), quiet-sun (QS), and open field (OF) regions

| Date | Region | Height [Mm] | 0 waves | 1 wave | 2 waves | 3 waves | ≥ 4 waves | Total threads | Total waves |
|---|---|---|---|---|---|---|---|---|---|
| 2010-05-23 | CH | 15.0 | 72 (15.4%) | 204 (43.7%) | 132 (28.3%) | 42 (9.0%) | 17 (3.6%) | 467 | 662 |
| 2010-05-23 | QS | 15.0 | 84 (18.2%) | 208 (45.1%) | 120 (26.0%) | 34 (7.4%) | 15 (3.3%) | 461 | 610 |
| 2010-10-06 | CH | 15.0 | 75 (15.0%) | 251 (50.2%) | 127 (25.4%) | 38 (7.6%) | 9 (1.8%) | 500 | 655 |
| 2012-03-27 | OF | 7.0 | 34 (6.7%) | 291 (57.1%) | 149 (29.2%) | 25 (4.9%) | 11 (2.2%) | 510 | 708 |
| 2012-03-27 | OF | 15.0 | 82 (15.4%) | 255 (47.9%) | 149 (28.0%) | 34 (6.4%) | 12 (2.3%) | 532 | 703 |

## (4.2) COMPARISON TO PREVIOUS STUDIES

We compare our NUWT results to three bulk measurements of transverse waves in polar plumes – one using indirect methods (McIntosh et al. 2011) and two using direct observations (Thurgood et al. 2014; Morton et al. 2015). McIntosh et al. (2011) studied a ~1 hour period of *SDO* / AIA 171 Å taken from 01:39–02:54 UT on 2010 April 25. They compared time-distance diagrams of the data to Monte Carlo simulations of transverse waves and concluded that the observations at 15 Mm were most similar to waves with velocity amplitudes of 25 ± 5 km s$^{-1}$ in the south polar coronal hole and 20 ± 5 km s$^{-1}$ in a quiet-sun region. For both regions, they also determined that the wave periods were within the range of 150 - 600 s (1.67 – 6.67 mHz).

The studies of both Thurgood et al. (2014) and Morton et al. (2015) also used *SDO* / AIA 171 Å data and utilized earlier versions of the NUWT code. Thurgood et al. analyzed five data slits ranging from 8.7 Mm to 34.8 Mm above the north polar coronal hole from 00:00–04:00 on 2010 August 6. They applied a Levenberg-Marquart fitting algorithm coupled with meticulous (and labor-intensive) user supervision to manually fit wave parameters to each identified thread. At a height of 15.2 Mm, they reported log-normal distributions of wave parameters with a mean amplitude of 498 ± 349 km, mean period of 200 ± 141 s (5 mHz), and mean velocity amplitude of 17 ± 12 km s$^{-1}$. Morton et al. used an earlier iteration of the FFT method and studied a single slit 6.96 Mm above the north polar coronal hole from 18:00–20:10 UT on 2012 March 27. They determined mean values of 591 ± 442 km, 414 ± 412 s (2.4 mHz), and 14.7 ± 15.6 km s$^{-1}$ for the displacement amplitude, period, and velocity amplitude respectively.

Figure 5 and Table 2 compare the wave parameters observed for each NUWT data slit to the values found in the aforementioned studies. The blue boxes span from the first (25%) to third (75%) quartiles of each set of waves. Solid lines within the boxes give the median parameter values, while red diamonds indicate the log-normal means. The box "whiskers" show the log-normal standard deviations for each variable. Complete sets of quartile values are unavailable for the previous studies, we instead plot the reported mean values with error bars, indicating the either the range of values (McIntosh et al.) or standard deviation (Thurgood et al. and Morton et al.). The NUWT extracted wave parameters exhibit a wide range of values that are reasonably well described by log-normal distributions with similar parameters for each slit. Within the IQR, we observe wave amplitudes in the range of 400 – 1000 km, periods of 150 – 500 s (frequencies of 2

– 6.7 mHz), and velocity amplitudes of 10 – 25 km s$^{-1}$. The corresponding log-normal means span, 688 – 836 km, 359 – 410 s (2.4 – 2.8 mHz), and 15.8 – 18.8 km s$^{-1}$. For the sake of comparing equivalent data, the values reported in Table 2 and Figure 5 have not been adjusted for rotation effects.

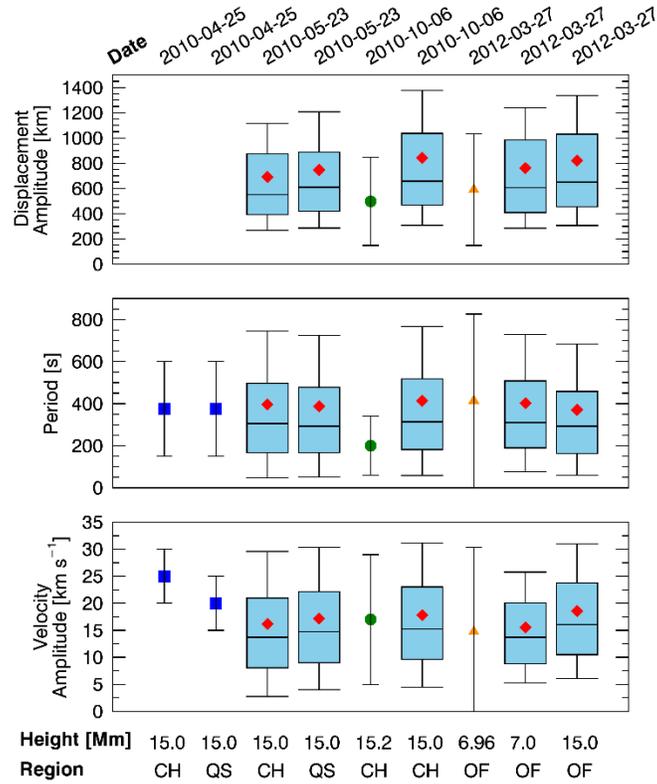

**Figure 5.** "Box and whisker" plot of the wave parameters within each NUWT data slit (blue boxes) as well as reference values from other studies. The lower and upper boundaries of the box indicate, respectively, the first (25%) and third (75%) quartiles. Horizontal lines within the boxes give the median values and the symbols give the log-normal means. The "whiskers" on the boxes show the log-normal standard deviations. The red diamonds indicate results found by our present study. The values marked with blue squares are from the paper by McIntosh et al. (2011). The results marked with a green circle are from Thurgood et al. (2014) and the orange triangle is from Morton et al. (2015). See also Table 2 for a comparison of the mean log-normal values.

Our results are in good agreement with the previously found values but reveal a much richer picture of the transverse wave behavior. We note that the FFT versions of NUWT (this work and that of Morton et al. 2015) tend to yield larger amplitudes and periods than the manual fitting version (Thurgood et al. 2014). However, the results are still within 1 standard deviation of each other and the differences may be a consequence of the finer control and quality of filtering (or user bias) afforded by the manual method. In all cases, the NUWT velocity amplitudes are smaller than those found by McIntosh et al. (2011). Additionally, we find little to no significant differences between the velocity amplitudes in the CH and QS regions, while McIntosh et al. noted a difference of 5 km s$^{-1}$.

**Table 2.** Comparison of wave parameters reported in coronal plumes

| Study | Date | Region | Height [Mm] | Amplitude [km] | Period [s] | Velocity amp. [km s$^{-1}$] |
|---|---|---|---|---|---|---|
| McIntosh et al. 2011 | 2010-04-25 | CH | 15 | -- | 150 - 600 | 25 ± 5 |
| " " | 2010-04-25 | QS | 15 | -- | 150 - 600 | 20 ± 5 |
| This work | 2010-05-23 | CH | 15.0 | 684 ± 425 | 409 ± 367 | 15.6 ± 13.0 |
| " " | 2010-05-23 | QS | 15.0 | 734 ± 453 | 398 ± 350 | 16.6 ± 13.0 |
| Thurgood et al. 2014 | 2010-10-06 | CH | 15.2 | 498 ± 349 | 200 ± 141 | 17 ± 12 |
| This work | 2010-10-06 | CH | 15.0 | 831 ± 532 | 421 ± 367 | 17.3 ± 13.0 |
| Morton et al. 2015 | 2012-03-27 | OF | 6.96 | 591 ± 442 | 414 ± 412 | 14.7 ± 15.6 |
| This work | 2012-03-27 | OF | 7.0 | 762 ± 480 | 407 ± 331 | 15.3 ± 10.1 |
| " " | 2012-03-27 | OF | 15.0 | 809 ± 504 | 378 ± 320 | 18.2 ± 12.5 |

Column headers: Mean Log-Normal Values (spanning Amplitude, Period, Velocity amp.)

## (4.3) RELATIONSHIP BETWEEN WAVE PERIODS

One of the key advantages of NUWT is the ability to identify multiple wave components superimposed within the same structure. As we found in § 4.1, between 34.8 – 40.9% of all threads detected by NUWT in the AIA data exhibited two or more significant waves. In contrast, only 8.6% of threads in the basic simulation and 11.4% of threads in the rotated simulation displayed multiple wave signatures. A majority (~2/3) of the simulated threads with multiple waves were the result of two or more input threads becoming merged into the same output thread. Within real data, the merging of threads corresponds to the cases where either the NUWT wave tracking method skips laterally to an adjacent structure or there is a change in the wave parameters observed within a single structure due to the propagation of multiple independent wave packets. Both cases will result in multiple signatures in the FFT spectrum, but the combined wave profile will have a generally poor fit to the data. A careful inspection of the AIA results indicates that, unlike the events in the synthetic data, the additional waves in the AIA data do not appear to be caused by the combination of different structures into the same output thread. Furthermore, the combined wave profiles display reasonably good fits to the peak intensity locations. Therefore, we conclude that the multiple wave results for the AIA data are the result of either the superposition of waves or multiple independent wave packets observed at different times. It is not possible to reliably distinguish between the two cases using only a single FFT. However, given the quality of the fits observed, the superposition of multiple waves appears to be more common. Further developments of the NUWT code may enable a more robust classification of FFT spectra and yield detailed statistics.

In each FFT spectra, we classify the wave with the largest amplitude as the "primary" wave and the wave with the second largest amplitude as the secondary wave. The magnitude of most secondary wave amplitudes is between 50 – 80% the magnitude of the primary wave amplitude. Therefore, these secondary waves may transport a significant portion of the total wave energy in the corona. In most cases, the primary wave also had the longest period (i.e. lower frequency); however, there were a number of events in which the secondary (or even tertiary) wave had a longer period than the primary.

Figure 6 shows histograms of the ratio between the first two significant wave periods within all NUWT threads with two or more waves. For simplicity, the ratio was calculated by dividing the longer period by the shorter period, regardless of how large the associated wave amplitudes were. Each color corresponds to one of the data slits examined and each histogram has been normalized by the total number of threads observed within that data slit. The results for the basic and rotated simulations are also shown plotted, respectively, in grey and gold. The thick black line represents the mean histogram for the five AIA slits. All AIA data slits exhibit a similar pattern. The ratios span a wide range of values with a notable peak between 2.4 – 3.2, a lesser peak (or at least a plateau) around 1.8 – 2.2, and the suggestion of another possible peak at 4.0 – 4.2. The QS slit on 2010-05-23 (light blue) has a larger proportion of values around 2.4 and the 7 Mm height slit on 2012-03-27 (violet) is the most sharply peaked. The histograms for the simulated data slits have more uniform distributions and display no statistically significant peaks.

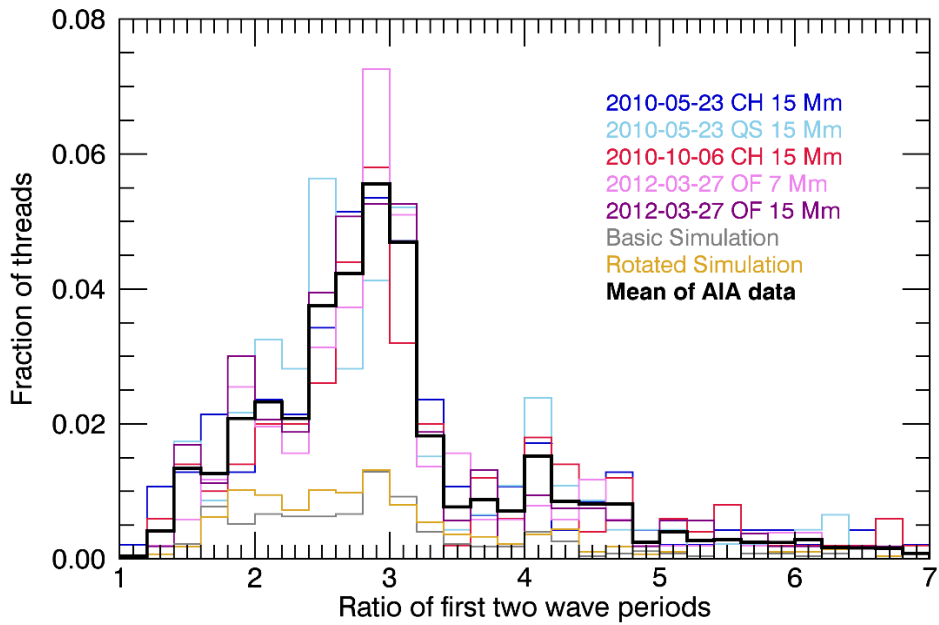

**Figure 6.** Histogram of the ratio between the first two significant wave periods within all NUWT threads with two or more waves. Each color corresponds to one of the data slits or simulations examined in this paper. The thick black line represents the mean histogram of the five AIA slits. All AIA data slits exhibit a similar pattern. The ratios span a wide range of values with a clear peak between 2.4 – 3.2 and minor peaks (or plateaus) around values of 2 and 4.

The cause of the peaks within Figure 6 is currently unknown. Harmonic relationships have been observed for standing mode waves within closed coronal loops (Verwichte et al. 2004; Van Doorsselaere et al. 2007). However, it is somewhat surprising to find similar behavior within an open field region. Preliminary explanations include wave interactions or a change in the frequency of the process driving the waves. At this time, observational effects and nonlinearities in peak intensity locations cannot be entirely excluded either. We can, however, conclude that the observation of multiple wave signatures is not simply an artifact of the FFT analysis methods, otherwise the simulated data would display a similar rate of occurrence. A full investigation concerning the exact nature of the relationship between waves within plumes is outside the scope of our present analysis.

## (5) SUMMARY AND CONCLUSIONS

We have described the development and validation of a fully automated version of the Northumbria University Wave Tracking (NUWT) code. Results from testing with synthetic data indicate, in the ideal case, that the returned mean sample amplitude is within 1-2% of the input population value while the mean period and velocity amplitude are accurate to within 4-9%. However, if the waves are rotated relative to the observational plane, the magnitude of the detected amplitudes will be reduced. Therefore, the amplitudes reported by NUWT represent, at worst, the lower bounds of the actual values. In the case of waves with uniformly distributed rotation angles, a scale factor of $\sqrt{2}$ can be applied to obtain more accurate bulk parameters. Better yet would be to combine NUWT observations with detailed information concerning the geometry of the structures that the waves are propagating within.

Using NUWT, we investigated the transverse wave motions within coronal plumes. We surveyed five, four-hour long data slits positioned above the solar limb: three located at a height of 15 Mm above a polar coronal hole, one at 15 Mm within a quiet-sun region, and a final coronal hole slit at 7 Mm. In total, NUWT detected 3,338 distinct waves within 2,470 separate features. The bulk wave parameters were found to be largely consistent with previous studies. Furthermore, between 34.8 – 40.9% of the observed features contained multiple waves at different frequencies. These additional waves may contain a non-negligible portion of the total wave energy.

Previous estimations for the total energy flux contained within the waves do not agree. Using an idealized equation, McIntosh et al. (2011) estimated energy flux densities on the order of $E_A = 100 - 200$ W m$^{-2}$ within both CH and QS regions at a height of 15 Mm. This is comparable to the energy required to accelerate the solar wind (Le Chat et al. 2012). However, using the same equation and range of coronal parameters, Thurgood et al. (2014) observed significantly less energy; only $E_A = 9 - 24$ W m$^{-2}$. Following the methods of these two studies, we find $E_A = 14 - 35$ W m$^{-2}$. If we assume the waves are randomly rotated with respect to the AIA imaging plane (see § 3.2), then the scaled velocity amplitudes yield an energy flux of $28 - 71$ W m$^{-2}$. While greater than the results of Thurgood et al., this is still less than the energy required to heat and accelerate the solar wind. Additionally, the above estimates use simplified equations that assume volume-filling waves in a homogeneous plasma. Models including more realistic filling factors and density profiles suggest that the total energy flux may be overestimated by factors of 5 – 10 (Van Doorsselaere et al. 2014) or even 10 – 50 (Goossens et al. 2013). Further research is needed to ascertain whether our observations are in some way incomplete or if other mechanisms and considerations are required. In particular, a more detailed and nuanced study of the full wave power spectra should be undertaken instead of relying on mean or median values that may fail to properly represent the shape of the parameter distributions.

The automated methods employed by NUWT provide a significant improvement to the speed at which we can directly measure transverse waves within the solar atmosphere. The code may also be applied to a wide range of imaging data and is ideally suited for studying periodic motions that display little to no damping, such as the decay-less oscillations observed in coronal loops (Anfinogentov et al. 2013; Nisticò et al. 2014; Anfinogentov et al. 2015). Using NUWT, the depth and breadth of transverse wave observations may now be expanded to scales that were previously infeasible due to the time and user-intensive effort required. Moreover, the code can be easily extended and new analysis methods may be quickly tested and compared to previous results. Potential extensions to NUWT include applying the motion magnification algorithm of Anfinogentov and Nakariakov (2016) to better resolve amplitudes smaller than 0.5 pixels, using the Lomb-Scargle periodograms (Lomb 1976; Scargle 1982) for unevenly sampled data, and using short-time Fourier transform techniques, such as Welch's method (Welch 1967) to investigate the time-evolution of individual waves. Possible avenues of future study include a long-term analysis spanning most of a solar cycle, fine-scale changes of wave parameters with height, the relationship between multiple waves within the same structure, total energy flux measurements, and investigations into waves within active region loops.

# ACKNOWLEDGEMENTS

This material is based upon work supported by the US Air Force Office of Scientific Research, Air Force Material Command, USAF under Award No. FA9550-16-1-0032. The authors acknowledge IDL support provided by the UK Science & Technology Facilities Council (STFC). RJM & JAM further acknowledge STFC support from grants ST/L006243/1 & ST/L006308/1. RJM acknowledges the support provided by the Leverhulme Trust.